\documentclass[fleqn,10pt]{wlscirep}
\usepackage[utf8]{inputenc}
\usepackage[T1]{fontenc}
\usepackage[framemethod=tikz]{mdframed}
\usepackage{lipsum}

\title{Organisational Social Influence on Directed Hierarchical Graphs, from Tyranny to Anarchy}

\author[1,*]{Charlie Pilgrim}
\author[2,4]{Weisi Guo}
\author[3,4]{Samuel Johnson}
\affil[1]{The University of Warwick, Centre for Doctoral Training in Mathematics for Real-World Systems, Coventry, CV4 7AL, UK}
\affil[2]{Cranfield University, Centre for Autonomous and Cyberphysical Systems, Cranfield, MK43 0AL, UK}
\affil[3]{The University of Birmingham, Mathematics Department, B15 2TT, UK}
\affil[4]{The Alan Turing Institute, London, NW1 2DB, UK}

\affil[*]{charlie.pilgrim@warwick.ac.uk}
\affil[+]{W.G. and S.J. conceived of the presented idea. C.P. developed the theory and performed the simulations.  W.G. and S.J. supervised throughout. C.P. took the lead in writing the manuscript with input and guidance from W.G..}

\begin{abstract}
Coordinated human behaviour takes place within a diverse range of social organisational structures, which can be thought of as power structures with "managers" who influence "subordinates". A change in policy in one part of the organisation can cause cascades throughout the structure, which may or may not be desirable. As organisations change in size, complexity and structure, the system dynamics also change. Here, we consider majority rule dynamics on organisations modelled as hierarchical directed graphs, where the directed edges indicate influence. We utilise a topological measure called the trophic incoherence parameter, q, which effectively gauges the stratification of power structure in an organisation. We show that this measure bounds regimes of behaviour. There is fast consensus at low q (e.g. tyranny), slow consensus at mid q (e.g. democracy), and no consensus at high q (e.g. anarchy). These regimes are investigated analytically, numerically and empirically with diverse case studies in the Roman Army, US Government, and a healthcare organisation. Our work demonstrates the usefulness of the trophic incoherence parameter when considering models of social influence dynamics, with widespread consequences in the design and analysis of organisations.
\end{abstract}
\begin{document}

\flushbottom
\maketitle
%
%
\thispagestyle{empty}

\section*{Introduction}

Social and political systems display different types of order and structure, with very different outcomes. Small-scale informal organisations might be skill or power based, whereas large-scale social systems often involve complex politics. Some form of formal or latent hierarchy is present in almost all social organisations. Traditional military organisations are perhaps the prototypical example of rigid hierarchy, with a very ordered structure allowing instructions to be quickly passed from top to bottom \cite{Sage2013JanRomans}. While these kinds of singular hierarchies abound historically \cite{Laloux2014FebOrgs}, in the 18th Century the influential treatise "The Spirit of the Laws" laid down a political theory that rejected hierarchical structure of government and called for a separation of powers \cite{Cohler1989Sep}, i.e. balancing power between multiple hierarchies. Influenced by this treatise and other enlightenment thinking, the US Constitution, which among other things dictates the structure of the US Government, prescribes a series of checks and balances with the explicit intention of preventing a singular tyrannical exploitation of power \cite{Beeman2010AugUSConan}. On the other end of the scale, political anarchy has been described as a rejection of any form of hierarchy \cite{Franks2018MarAnarchy}. Anarchy should not be dismissed as disorganised chaos - there are movements in management \cite{Laloux2014FebOrgs} and organisational \cite{Buurtzorg} science that encourage self-empowered individuals working autonomously or in dynamically forming teams, with an organisational structure resembling that of political anarchy. One can regard anarchy as a non-equilibrium form of hierarchy, whereby at any particular quasi-static state, a power hierarchy exists. In between these extremes, modern democracies (on average) can be considered more distributed than rigid tyrannical hierarchies and more ordered than political anarchy, somewhere between the two. In fact, we can find a myriad of organisational structures \cite{Laloux2014FebOrgs}, ranging from the structured (military, churches, schools), to those that are consumer driven, to those that have an ad-hoc agile decision process. An important question in all of these social structures is how effective leadership is at influencing change, or how fast a change in policy spreads through the network, if at all. The same question sheds light on how fast a disruption can spread through an organisation, with resilient ones able to dampen its propagation quickly, whereas non-resilient organisations can suffer long-term cascading confusion.

Social systems are part of a broader class of complex interaction systems that are well suited to being modelled as networks. There has been much work in the past investigating the structure and dynamics of complex networks, with many excellent overviews including \cite{Boccaletti2006Feb, Newman2003Mar, Albert2001Jun}. Dynamics such as stability \cite{ParadoxMay1972Aug, Johnson2014Dec}, consensus \cite{Olfati-Saber2004SepConsensusConnectivity}, disease spread \cite{Shirley2005Sep} and percolation \cite{Cohen2009} have been examined in relation to topological measures such as spectral analysis \cite{ParadoxMay1972Aug, Boccaletti2006Feb}, connectivity \cite{Olfati-Saber2004SepConsensusConnectivity}, clustering \cite{Boccaletti2006Feb}, core-periphery \cite{Guo16}, degree distributions \cite{Boccaletti2006Feb} and trophic coherence \cite{Pagani19RSOS}, among many other studies. Joint entity dynamics and topological analysis have been conducted in recent studies, especially for one-dimensional population dynamics \cite{Barzel13, Gao16}.

There is a burgeoning recognition that there are a class of systems, and associated network models, where the direction of edges is of paramount importance to the system dynamics. Non-normality \cite{Asllani2017Jun} and trophic coherence \cite{Johnson2014Dec} have emerged as two important topological measures of directed graphs that address this. Non-normality can be thought of as the degree of departure from symmetry in a network, while trophic coherence is a measure of how aligned the edges are in one direction. There is a conceptual similarity, and the two measures have been proven to be mathematically connected so that networks that have high trophic coherence are also highly non-normal \cite{Johnson2019Aug}. These measures have had successes in describing systems where the direction of interactions is important, including  ecosystem resilience \cite{Johnson2014Dec, Asllani2017Jun} and neural dynamics \cite{Klaise2016JunGPPM, Asllani2018Dec}. In the social realm, we consider that the direction of edges (and influence) in social organisational models is important and so is suitable for this kind of analysis.

Social systems display rich and complex behaviour, far beyond our ability to fully capture in mathematical models. In order to investigate the problem we necessarily make simplifying assumptions. Our approach is informed by sociophysics, the application of models and methods from the physical sciences to social systems \cite{sociophysicsCastellano2007Oct, Guo16}. The most common and valid criticism of sociophysics is that the necessary assumptions made in order to reduce the social system to a tractable model can often overlook the subtleties and richness of the social dynamics \cite{sociophysicsCastellano2007Oct}. A response to this criticism, and the stance that we aim for, is to avoid the positivist pitfall of thinking that our models completely describe or predict social behaviour, and instead to look for simple models that reveal salient features of social systems, that can then be used to inform further research and more qualitative approaches. With this in mind, we consider networks with nodes as people and political influence moving along directed edges. We are interested in the dynamics on these networks, in particular the speed to consensus following a change in policy from the leadership. 
 
Our approach, which will be fully explained in the following section, is along the lines of previous work on modelling opinion dynamics in social systems by considering the dynamics of the Ising model in complex networks (see \cite{sociophysicsCastellano2007Oct} for an overview). Our model can lead to nodes flipping back and forth between states as influence spreads, which is an important aspect of social influence dynamics in the real world. This phenomena emerges from the propagation of multiple states (types of influence) through cycle structures that form in non-tree like directed graphs, which cannot easily be described using similar models such as the SI model in epidemiology \cite{Brauer2008} or cascade failure models \cite{Motter2002Dec}.

Our unique contribution will be to show how time to consensus on these networks changes in relation to the topological measure of the trophic incoherence parameter. The broader aim with this study is to show, through this simple example, how the consideration of trophic coherence can be applied to improving our understanding of social networks.

\section*{Methods}

We will model social systems as directed graphs, with nodes as people (or small groups/entities within organisations) and with influence spreading along directed edges. Here, influence is an asymmetric property (can be two-ways) that can be the delegation of a task, the transmission of a command, or the sharing of information. We define a directed graph as $\mathcal{G}=(V, E)$, where $V$ is the set of vertices or nodes in the graph, and $E$ is the set of directed edges between nodes. This can be fully described by an adjacency matrix $A$, such that the elements of $A$ are 

$$a_{ij} = \begin{cases} 1 &\text{if directed edge exists from } i \text{ to } j \\
0 &  \text{otherwise} \end{cases}
$$

We will use $S$ or $n$ to denote the size, or number of nodes, of a network. $L$ will be used for the number of edges. $B$ will be used for the number of basal nodes, which are nodes with no in-edges. From a social systems perspective, basal nodes are considered "leaders" who are not influenced by any other nodes. 

\subsection*{Power Stratification Measured by Topological Trophic Coherence}

In different social organisation structures, the clarity or coherence of power stratification depends on the number of feedback loops subordinates have to managers.
In our abstract model, if a subordinate has equal feedback power as the manager has delegate power, we regard them as equals. When the network is large, the feedback can arch across different levels and it becomes difficult to quantify: (1) the number of power levels, and (2) the coherence of the power levels. To capture the degree of power stratification we will consider trophic coherence, a topological property of directed networks which captures the extent to which edges are aligned, as in a top-down hierarchy, or more disorganised, as usually seen in a random graph \cite{Johnson2014Dec, LooplessnessJohnson2017May}. To describe trophic coherence, we first need to describe trophic levels.

The concept of trophic levels was originally developed in relation to food web networks, which classify species in accordance to their predation relationships \cite{Johnson2014Dec}. In a food web, the nodes are species and directed edges go from prey to predator. Each species can be assigned a trophic level, which signifies how far "up" the food web the species is \cite{Levine1980MarTrophicLevels}. Basal species, with trophic level 1, are those that generate energy directly from the environment such as plants and algae \cite{Levine1980MarTrophicLevels}. Species that feed only on basal species, such as sheep, are assigned trophic level 2. A wolf that predates only on sheep would be given trophic level 3. Some species do not fit neatly into an integer trophic level, for example a scavenger like a rat may feed on plants as well as the dead bodies of sheep and wolves \cite{Johnson2014Dec}. The concept of describing directed networks with energy flow has been extended beyond food webs, for instance to inferring multi-scale stability in both transportation networks \cite{Pagani19RSOS} and water distribution systems \cite{Pagani19ASCE}.

Consider the food web network adjacency matrix, $A$, such that $a_{ij}$ is the amount of biomass that species $j$ predates from species $i$. The trophic level of a species, $s_j$, is defined as the weighted mean of the trophic levels of the species that $j$ preys upon, plus one \cite{Johnson2014Dec}.

\begin{equation}
s_j = \frac{\sum_{i} a_{ij} s_i}{\sum_{i} a_{ij}} + 1 \label{eqn:trophic_level_definition}
\end{equation}

This describes a linear system that can be solved with a unique solution with the sufficient condition that each node has at least one path from a basal node to itself. A full derivation is in the Supplementary Information.

We define the trophic incoherence parameter, $q$, along the lines of Johnson et al (2014) \cite{Johnson2014Dec}. The trophic difference of an edge in the graph is the difference in trophic levels between predator and prey. 

\begin{equation}
d_{ij} = s_j - s_i
\end{equation}

The mean of the trophic difference of edges in any directed graph will be equal to 1 (see \cite{Moutsinas2019Aug} for a proof). In fact, in a perfectly ordered graph, the trophic difference of every edge will be 1 \cite{Johnson2014Dec}. In less ordered graphs, the mean of the trophic differences of edges will remain equal to 1, but with some variance. The trophic incoherence parameter, $q$, is defined as the standard deviation of the distribution of trophic differences over all edges in the graph \cite{Johnson2014Dec}. Conceptually, directed graphs that have high trophic coherence (i.e. low trophic incoherence parameter) are tree like, and can be drawn with all edges pointing in the same direction. Directed graphs with low trophic coherence (i.e. high trophic incoherence parameter) do not have edges pointing in one clear direction, and appear more random.

\begin{figure}
\centering
\includegraphics[width=1\linewidth]{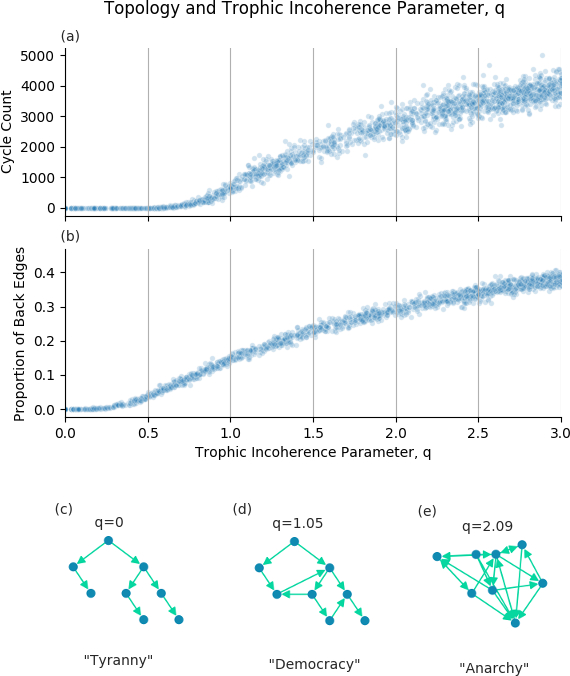}
\caption{The trophic incoherence parameter's relation to topological structures. $\mathbf{(a)}$ and $\mathbf{(b)}$ show counts of cycles of length 5 and the proportion of back edges in graphs generated using the general Preferential Preying Model with 100 nodes, 750 edges and 5 basal nodes, over a range of the trophic incoherence parameter. At low q $\mathbf{(c)}$, graphs are tree-like, analogous to a tyrannical social system. At high q $\mathbf{(e)}$, graphs are random with little coherent structure, analogous to anarchy. At mid q  $\mathbf{(d)}$, there is some coherence, analogous to democracy.}
\label{fig:trophic_incoherence}
\end{figure}

The concept of trophic coherence was initially introduced as a way to potentially solve May's paradox, a long standing problem in ecology \cite{Johnson2014Dec}. As a random graph grows in size and connectivity, it reaches a threshold beyond which it is almost certainly unstable (considering the Lyapunov stability of the community matrix) \cite{ParadoxMay1972Aug}. The experience of ecologists suggests that nature seems to behave the opposite way - ecosystems do not become less stable with size and complexity and can become more stable \cite{Johnson2014Dec, Jacquet2016Aug}. Light was shed on this apparent paradox in 2014 by Johnson et al, who showed that trophic coherence determines the stability of food-webs \cite{Johnson2014Dec}. Since then, trophic coherence has been associated with the presence of cycles in networks \cite{LooplessnessJohnson2017May}, and has been linked to the distribution of motifs \cite{MotifsKlaise2017Nov}. 

Trophic coherence has been called a measure of how similar a graph is to a hierarchy \cite{Pagani19RSOS}, with a hierarchy being maximally coherent with trophic incoherence parameter, $q=0$ (Figure \ref{fig:trophic_incoherence}(c)). In a hierarchy, all nodes are in a clear trophic level with edges all pointing from top to bottom. Conversely, graphs with a very high trophic incoherence parameter will appear random, with no clear structure of nodes and edges going in all directions (Figure \ref{fig:trophic_incoherence}(e)). Between these two extremes we have graphs which have a discernible order but are not fully hierarchical, with some edges going against the flow (Figure \ref{fig:trophic_incoherence}(d)). 

Figure \ref{fig:trophic_incoherence} shows relationships between the trophic incoherence parameter and topological structures. Johnson et al showed a theoretical relationship between cycles and trophic coherence \cite{LooplessnessJohnson2017May}, which suggested that graphs in the low q regime will have very few cycles, this is demonstrated empirically in \ref{fig:trophic_incoherence}(a). A discussion of how the number of cycles in a graph was calculated is included in the Supplementary Information.

A "back edge" is a link from a node of higher trophic level to a node of lower trophic level i.e. the trophic difference of this edge will be below zero. As described above, we know that the mean of the trophic differences of edges will be 1 for any graph, and that the trophic incoherence parameter is the standard deviation of the distribution of trophic differences. The probability of an edge having trophic difference below zero should therefore increase with the trophic incoherence parameter, $q$. Figure \ref{fig:trophic_incoherence}(b) shows that, empirically, the expected number of back edges appears to be an increasing function of the trophic incoherence parameter, $q$, as expected.

The empirical number of cycles and back edges support our conceptual view of trophic coherence. This gives us 3 regimes of structure, each of which has a social analogus:

\begin{enumerate}
  \item Low q regime. Hierarchical, tree-like graphs with few cycles or back edges. The social analogue is Tyranny. 
  \item Medium q regime. Broadly hierarchical, but with increasing numbers of cycles and back edges. The social analogue is Democracy.
  \item High q regime. Little hierarchical structure, lots of cycles and back edges. The social analogue is Anarchy.
\end{enumerate}

For our purposes, trophic coherence is a useful concept to describe different types of social systems, and in particular how coherent a social structure is. The trophic incoherence parameter, $q$, gives us a one-dimensional measure of the slightly abstract concept of coherence, which will allow us to investigate the structure of social systems in a novel and quantitative way.

\subsection*{Network Generation Model}

One of the earliest and most influential random graph generation models is the Erd\H{o}s-{}R\'enyi model \cite{Bollobas1998}. The Erd\H{o}s-{}R\'enyi graph generating algorithm with node count $S$ and edge count $L$ uniformly randomly places $L$ directed edges across all possible edges between nodes \cite{erdos59a}. The ensemble includes all possible configurations of $\mathcal{G}=(V,E)$ with those node and edge counts \cite{erdos59a}. However, Erd\H{o}s-{}R\'enyi graphs have a very low probability of having high trophic coherence. While it is theoretically possible to investigate the entire state space with Erd\H{o}s-{}R\'enyi graphs, we seek a more reliable way of tuning random graphs with specific trophic incoherence parameters, $q$.

The niche and cascade models arose from the literature on ecology, as attempts to generate artificial food webs similar to empirical food webs using simple rules \cite{Johnson2014Dec}. These models generate directed graphs with trophic incoherence parameters in the range $0.5 \lessapprox q \lessapprox 2$. However, we need to generate graphs with a wider range of trophic coherence. 

Johnson et al's 2014 paper, which introduced the trophic incoherence parameter, $q$, also introduced the Preferential Preying Model \cite{Johnson2014Dec}. This was updated in 2016 as the general Preferential Preying Model \cite{Klaise2016JunGPPM}, which we will use as our main graph generating model. The Preferential Preying Model was inspired by considering how ecological food webs form in nature \cite{Moutsinas2019Aug}, and it has the advantage of generating graphs over a controllable range of trophic coherence. We will use this graph generation model during our investigation. 

The general Preferential Preying Model algorithm requires an input of the number of nodes, $S$, the number of basal nodes, $B$, the number of links, $L$ and a Temperature parameter, $T$, which is used to tune the trophic incoherence parameter of the resulting graph \cite{Klaise2016JunGPPM}. The algorithm proceeds in 2 phases.

The algorithm begins with $B$ basal nodes, which are given a temporary trophic level, $\hat{s}_i=1$. Each of the remaining $S-B$ nodes are added one at a time, with an edge created to the new node, $j$, from an existing node, $i$, which is chosen uniformly randomly. The temporary trophic level of the new node is set to $\hat{s}_j=\hat{s}_i + 1$. Once all nodes are added we have a graph with $S$ nodes and $(S-B)$ edges \cite{Klaise2016JunGPPM}. 
In phase two, $L-(S-B)$ more links are added, the relative probability of a link being added from node $i$ to node $j$ is given by:

\begin{equation}
p_{ij} \propto exp \left(-\frac{(\hat{s}_j-\hat{s}_i -1)^2}{2T^2} \right)
\end{equation}

The general Preferential Preying Model generates graphs from a particular ensemble \cite{LooplessnessJohnson2017May}. Real social systems may exist outside of this ensemble, and may have dynamic structures that change over time. It may be that there is some hidden feature of graphs generated using the general Preferential Preying Model which is responsible for any results or conclusions we find. We will consider graphs generated using the general Preferential Preying Model as a null model, and be wary of over generalising our results to all social systems. 

\subsection*{Opinion Dynamics}

The majority rule model \cite{sociophysicsCastellano2007Oct} allows each node to be in one of two states, which we will denote as state $-1$ or state $1$, $x_i \in \{-1, 1\}$. We define the state vector as the state of all $S$ nodes, $\mathbf{x} = [x_1, x_2, ... ,x_n]$.  We consider nodes being influenced along the directed edges, with the state of a node changing based on some form of influence dynamics. Considering discrete timesteps, the state of nodes in a timestep are determined by some function on the graph topology and the state of the nodes in the previous timestep \cite{sociophysicsCastellano2007Oct}. 

\begin{equation}
\mathbf{x_t} = f(\mathcal{G}, \mathbf{x_{t-1}})
\end{equation}

We will use a simple majority rule algorithm to update the states of the nodes at each timestep. At each timestep, $t$, a node is uniformly randomly selected and its state is updated based on the state of its parent nodes, where a directed link goes from a parent to a child node.  The algorithm will update a node's state to match the state of the majority of its parent nodes in the previous timestep \cite{sociophysicsCastellano2007Oct}. If there is a tie then the node's state will be determined by a coin flip, with a $50 \% $ probability of state $-1$ or $1$ \cite{sociophysicsCastellano2007Oct}.

\section*{Analysis}

\begin{figure}
    \centering
    \includegraphics[width=1\linewidth]{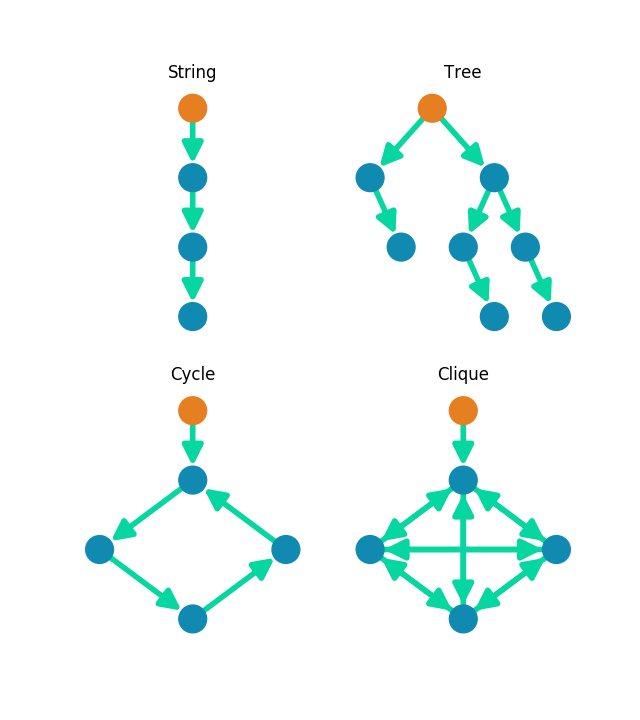}
    \caption{Selected types of simple graphs for analysis. Orange nodes are basal nodes, which will begin in a different state to the non-basal nodes.}
    \label{fig:motifs}
\end{figure}

We will analytically investigate how we expect majority rule dynamics to propagate on a series of simplified models, which we can then compare to empirical results. We will look at 4 types of simple graphs, as shown in Figure \ref{fig:motifs}.

\begin{enumerate}
  \item Strings.
  \item Trees. 
  \item Cycles. 
  \item Cliques. 
\end{enumerate}

For each type of simple graph, we will consider initialising the system so that the basal node is in state -1 and all non-basal nodes are in state 1. We will find expressions for $v(t)$, the number of nodes in state 1 (not yet influenced by the basal node) as a function of timesteps, $t$.

\subsection*{Strings}

The simplest form of a graph with high trophic coherence (and low trophic incoherence parameter, $q$) is a string, a line of nodes with edges down the line in one direction. We can initialise this string so that all nodes are in state 1 except the basal node, in state -1 (Figure \ref{fig:motifs}). We let $v(t)$ be the number of nodes in state 1 as a function of timesteps. Each timestep can be considered a Bernoulli trial, with probability of success equal to the probability of the algorithm selecting the next node in the string that will spread the influence of the basal node, $p=\frac{1}{n}$. Success in this Bernoulli trial will reduce $v(t)$ by 1, with $v(0)=(n-1)$. The sum of Bernoulli trails gives a Binomial distribution \cite{Grinstead2012Oct}, so that we can write

\begin{equation}
(n-1) - v(t) \sim Binomial(t, \frac{1}{n}) \quad \quad v \in \{0, 1, .., n-1\}
\end{equation}

and the expected value of $v(t)$ can be easily found and written as

\begin{equation}
<v(t)>=
\begin{cases}
n-1-\frac{t}{n} &\mbox{if } t \leq n(n-1)\\
0  &\mbox{if } t > n(n-1)  \\
\end{cases}
\end{equation}

\begin{figure}
    \centering
    \includegraphics[width=1\linewidth]{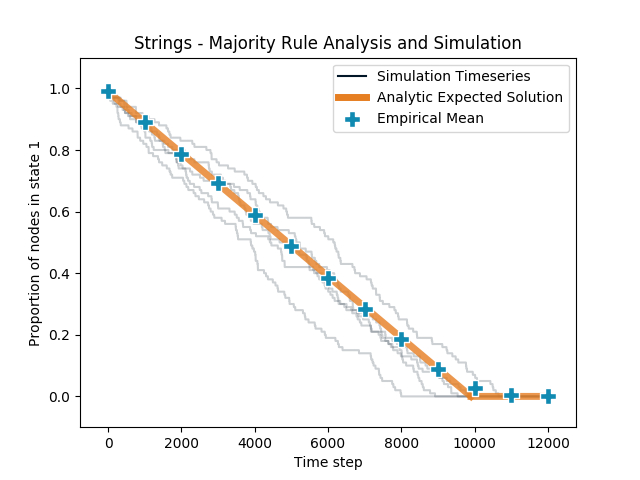}
    \caption{Progress of influence dynamics on a directed string of length 100, with one basal node beginning in state -1 and all non-basal nodes beginning in state 1. The orange line shows the analytic expected proportion of nodes in state 1 as a function of timesteps. The blue points show the empirical mean of 50 simulated time series at 1000 timestep intervals. The light grey lines are 5 of those simulated time series.}
    \label{fig:motifFixationString}
\end{figure}

Figure \ref{fig:motifFixationString} shows this expected value against simulations, with close agreement. 

\subsection*{Trees}
We define a tree as a directed graph with no cycles, and with directed edges going in one direction only from basal nodes to leaf nodes, where a leaf node is a node with only in-edges. Again, we initialise the tree with basal nodes in state -1 and non-basal nodes in state 1 (Figure \ref{fig:motifs}). We are interested in $v(t)$, the number of nodes in state 1 as a function of timesteps under simple majority rule dynamics. 

In a tree, the number of nodes influenced by the basal node can only increase, as all edges go from the basal node outwards. As the basal node's influence spreads, there is a "fringe" of nodes that have yet to be influenced by the basal node, but who have a chance of switching in the next timestep if they are chosen. This fringe, and so the probability of spreading influence per timestep, changes over the  progress of the simulation. The spread of influence can be modelled as a series of Bernoulli trials with a varying probability of success, $p_{tree}$. If we assume that $p_{tree}$ is Beta distributed, then the number of nodes influenced by the basal node as a function of time will be Beta Binomial distributed \cite{Johnson2005SepBetaNegBinomial}, with the form of $p_{tree}$ determined by the Beta distribution parameters $\alpha$ and $\beta$, which will depend on the topology of the tree in question and the progress of the simulation. 

\begin{equation}
(n-1) - v(t) \sim Binomial(t, p_{tree}) \quad \quad v \in \{0, 1, .., n-1\}
\end{equation}

where $p_{tree}$ is a Beta distributed random variable

\begin{equation}
p_{tree} \sim Beta(\alpha, \beta)
\end{equation}

The number of timesteps to consensus can be modelled by considering the number of failed Bernoulli trials before $0.9n-B$ successes (consensus requires 90\% of nodes to agree with the basal nodes). In a network of 100 nodes and 5 basal nodes, consensus is reached when 90 nodes agree with the basal nodes, which requires 85 successful Bernoulli trials, or 85 successful algorithm timesteps where a node is switched state to agree with the basal nodes. This kind of process with a fixed $p$ will have a Negative Binomial distribution \cite{Johnson2005SepBetaNegBinomial}. As discussed, $p$ is not fixed for trees. If we approximate $p_{tree}$ as being a Beta distributed random variable, then the time to consensus can be approximated as a Beta Negative Binomial distribution \cite{Johnson2005SepBetaNegBinomial}. We expect this distribution for timesteps to consensus for tree-like graphs, this will be investigated empirically in the Results section.

\subsection*{Cycles}
Cycles are loops of nodes with directed links pointing around the loop in one direction. We will consider a cycle of nodes with one basal node outside the cycle, influencing one node in the cycle (Figure \ref{fig:motifs}). The basal node is always in state -1. Intuitively, it can be seen that the rate at which the basal node influences the cycle is proportional to the number of nodes in the cycle in state 1, yet to be influenced. This leads to an exponential relationship for the expected value of $v(t)$, a full derivation is given in the Supplementary Information. 

\begin{equation}
    <v(t)> = (n-1)e^{-\frac{ln2}{n(n-1)}t}
\end{equation}

\begin{figure}
    \centering
    \includegraphics[width=1\linewidth]{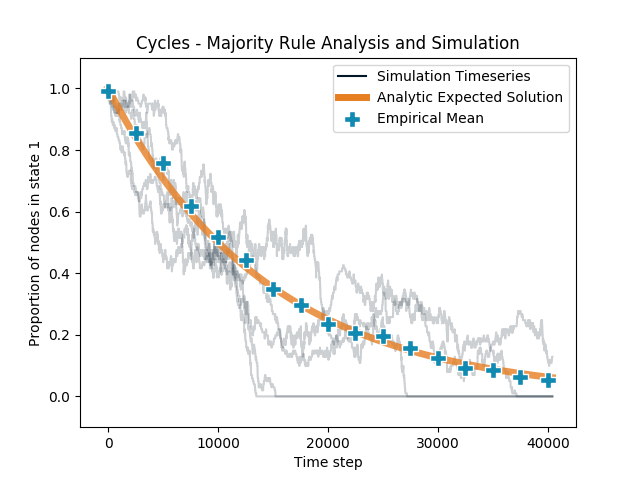}
    \caption{Progress of influence dynamics of a directed cycle of length 100, with one additional basal node beginning in state -1 and all non-basal nodes beginning in state 1. The orange line shows the analytic expected proportion of nodes in state 1 as a function of timesteps. The blue points show the empirical mean of 50 simulated time series at 2500 timestep intervals. The light grey lines show 5 of those simulated time series.}
    \label{fig:motifFixationCycle}
\end{figure}

Figure \ref{fig:motifFixationCycle} shows this analytic prediction against simulated results for simple majority rule dynamics. The simulated results appear to be randomly oscillating but with an expected value over many simulations close to the analytic prediction. In the simulations, the random oscillations of $v(t)$ mean that the system can reach the absorbing state of $v=0$. The analytic expected value is in the limit of large n, where the probability of the system reaching the absorbing state is small. 

\subsection*{Cliques}
We define a clique in a directed graph as a group of nodes such that edges link each node to all other nodes in the group, in both directions. We consider one basal node in state -1 influencing one node within the clique, all of which are in state 1 (Figure \ref{fig:motifs}). In a clique above size 2 each node reinforces the state of the other nodes, and a single basal node is unable to influence any of the nodes in the clique, which all therefore stay in their initial state of 1. There is no possibility of consensus with the basal node. These kinds of structures within graphs can block any influence spreading from the basal nodes. $v(t)$ will stay at its initial value of $n-1$.

\begin{equation}
v(t) = n-1
\end{equation}

\section*{Results}

\subsection*{Time to Consensus}

We are interested in how quickly a network will reach consensus following a change in policy from the leadership. We consider here basal nodes, with no parent nodes (no in-edges), as "leaders". We initialise the state of the basal nodes to be $x_{basal}=-1$ and the non-basal nodes to be $x_{non-basal}=1$. We then run a simulation based on majority rule dynamics, and record how many timesteps it takes for $90 \%$ of the nodes to agree with the state of the basal nodes, which we consider to be consensus. If consensus is not reached before a set number of timesteps then the simulation is stopped. We repeat this for graphs generated using the general Preferential Preying Model, with a range of trophic coherence.

\begin{figure*}
\centering
\includegraphics[width=1\linewidth]{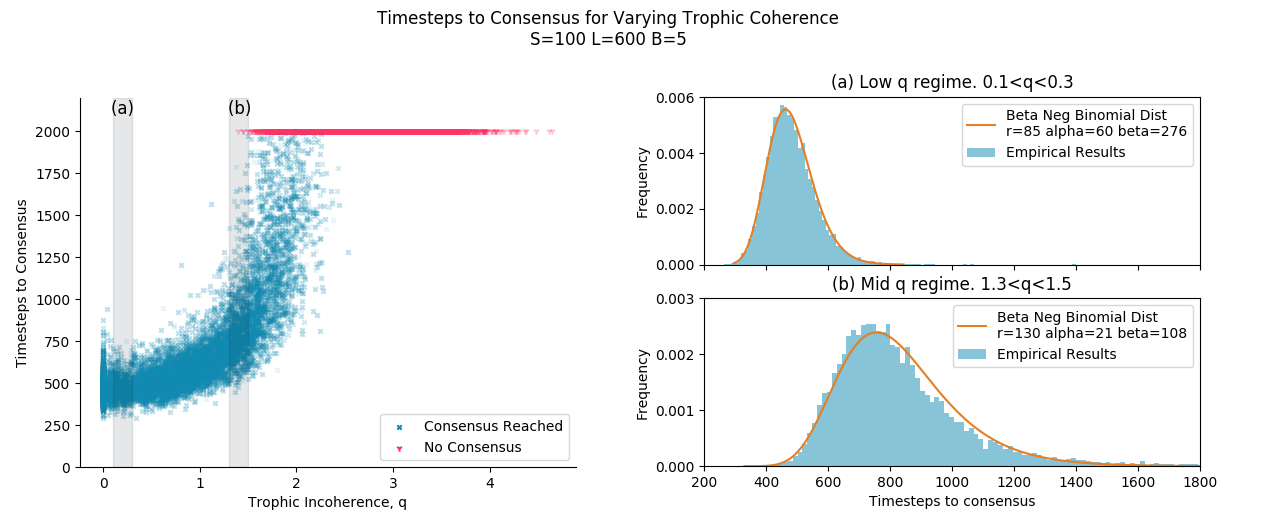}
\caption{The grey highlighted sections in the left figure correspond to the figures on the right, marked $\mathbf{(a)}$ and $\mathbf{(b)}$. The left figure shows the result of simulations of majority rule dynamics on graphs generated using the general Preferential Preying Model, with each simulation representing one point. Simulations that did not reach consensus after 2000 timesteps are marked at 2000 timesteps in red. $\mathbf{(a)}$ shows the distribution of time to consensus for graphs generated in the range between $0.1<q<0.3$, indicated by the shaded region in the left figure. The Beta Negative Binomial distribution is a good fit with $r=85$ and $\alpha$ and $\beta$ adjusted manually, with the constraint that the empirical mean and distribution mean remain equal. $\mathbf{(b)}$ shows the distribution of time to consensus in the range $1.3<q<1.5$ (for simulations that reached consensus). Graphs take longer to reach consensus at this higher q range, and so the entire empirical distribution is shifted to the right. The Beta Negative Binomial distribution is not as good a fit as in the low $q$ regime, and the $r$, $\alpha$ and $\beta$ parameters were all adjusted manually.}
\label{fig:time_to_consensus}
\end{figure*}

Figure \ref{fig:time_to_consensus} shows the time to consensus for graphs generated with a range of trophic coherence and a fixed number of nodes, $S$, edges, $L$, and basal nodes, $B$. This kind of pattern is consistent across a range of values for S, B and L. 

From a visual inspection of Figure \ref{fig:time_to_consensus}, it can be seen that for graphs with a low trophic incoherence parameter, $q\lessapprox 1$, there is very little change in time to consensus with changing $q$. As $q$ increases above 1, we increasingly see graphs with longer times to consensus. At around $q = 1.5$, we start seeing graphs take much longer to reach consensus, and some that do not reach consensus at all. In the results here, simulations that ran for 2000 timesteps were considered as not reaching consensus. When simulations were ran for 5000 timesteps instead, there were some that reached consensus after 2000 timesteps in the $1.5 \lessapprox q \lessapprox 2.5$ range. The cutoff of 2000 timesteps in Figure \ref{fig:time_to_consensus} was chosen to clearly and faithfully show the structure of the data over a range of the trophic incoherence parameter. 

This gives us 3 regimes of behaviour to describe:

\begin{enumerate}
  \item Fast consensus. Low q. Policy changes spread through the network very fast. 
  \item Slow consensus. Medium q. Policy changes eventually spread through the network. 
  \item No consensus. High q. Policy changes do not spread throughout the network.
\end{enumerate}

These regimes are approximate, and are not fully captured by the trophic incoherence parameter. For example in Figure \ref{fig:time_to_consensus} for trophic incoherence parameter $1.5 \leq q \leq 2$, some graphs reach consensus quite quickly, some take a long time, and some do not reach consensus at all. The regimes of behaviour do seem to be bounded by the trophic incoherence parameter, with clear regions for each regime, although with some overlap. An obvious question is what is it about the topology of the graphs in the different regimes that is causing the different behaviour in influence dynamics. 

\subsubsection*{Fast Consensus Regime, "Tyranny" - Low q }

At low $q$, we see graphs with fast consensus. The graphs here should all be tree-like, and conceptually it is expected that they should reach consensus quickly. Figure \ref{fig:time_to_consensus}(a) shows a reasonably good fit of the Beta Negative Binomial distribution to the empirical distribution of time steps to consensus of an ensemble of graphs with trophic incoherence parameter $0.1<q<0.3$. This matches what we predicted in the Analysis section. The Beta Negative Binomial distribution here was fit with the number of successful trials, $r=0.9S-B=85$. The Beta distribution shape parameters of $\alpha$=60 and $\beta=276$ were adjusted manually. These values give quite a narrow Beta distributed $p_{tree}$ value, with $<p_{tree}>=0.18$.

\subsubsection*{Slow Consensus Regime, "Democracy" - Medium q}

In the medium q range the dynamics of consensus change. Consensus is reached, but it takes longer. From Figure \ref{fig:trophic_incoherence} we can see that cycles begin to appear in large numbers in networks with trophic incoherence parameter above around $q=1$, and back edges appear from about $q=0.5$. The presence of cycles and back edges mean that $v(t)$ is not necessarily a strictly decreasing function. During the simulation, nodes can change states to disagree with the basal nodes, so that $v(t)$ can go up, stay the same or go down. Each timestep is no longer a Bernoulli trial, as there are 3 possible outcomes. Conceptually, it makes sense that consensus takes longer in these kinds of graphs. Despite the fact that the underlying assumptions of Bernoulli trials no longer hold, the Beta Negative Binomial distribution is still a reasonable fit to the empirical data, as shown in Figure \ref{fig:time_to_consensus}(b). Consensus takes longer than in the low q regime, with a wider distribution, and not as good a fit. The number of successful trials, $r=130$, to consensus was not predetermined during the distribution fit, because if some nodes switch against the basal nodes state then it will take more than 85 successes to reach consensus. $\alpha=21$ and $\beta=108$ give a wider Beta distribution for the value of $p_{tree}$ than for the low q regime, with $<p_{tree}>=0.16$.

\subsubsection*{No Consensus Regime, "Anarchy" - High q}

As the trophic incoherence parameter rises above around $q=1.5$, we start to see some graphs that do not reach consensus at all. As the trophic incoherence parameter rises further above around $q=2.5$, all graphs are in the no consensus regime. The basal nodes, or leaders, are unable to influence the rest of the nodes. At these high q values, we expect to see a high number of cycles and back edges (Figure \ref{fig:trophic_incoherence}). We hypothesise that the dynamics are similar to those found in the cliques in the Analysis section. This may be caused by nodes with a high number of links from nodes of a higher trophic level, which act to block influence spreading from the basal nodes, preventing consensus.

\subsection*{Average Degree and Network Size}

We have shown that majority rule dynamics, and time to consensus, are linked to the trophic coherence of a graph. In our analysis so far, we have concentrated on networks with 100 nodes, 600 edges and 5 basal nodes. It makes sense to begin the investigation by keeping everything constant except for the trophic incoherence parameter, however the question arises of whether similar results are found if we vary the average degree, $L/S$, and the size of the networks, $S$. 

\begin{figure}
    \centering
    \includegraphics[width=1\linewidth]{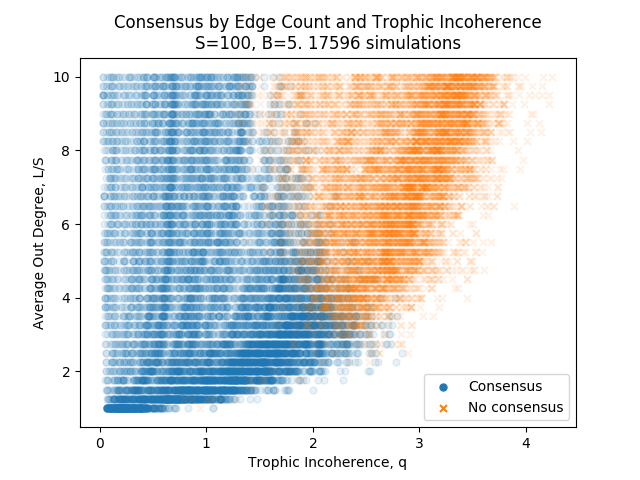}
    \caption{Each point represents a simulation of majority rule dynamics on graphs generated using the general Preferential Preying Model, either reaching consensus or not. All networks have 100 nodes, 5 basal nodes and varying average degree, $L/S$, and varying trophic incoherence parameter, $q$.}
    \label{fig:consensus_regimes_edges}
\end{figure}

Figure \ref{fig:consensus_regimes_edges} shows the regimes of consensus and no consensus for simulations on graphs with fixed node count and basal node count, with varying average degree, $L/S$ and trophic incoherence parameter, $q$. There are clear areas of consensus and no consensus, with some overlap. With higher average degree, the no consensus regime begins at lower values of the trophic incoherence parameter. A hypothesis is that more edges mean that it is more probable to find situations where nodes self-reinforce each other and block influence spreading, as seen in cliques in the Analysis section. The general Preferential Preying Model is unable to generate graphs with high trophic incoherence parameters and low edge counts, so the region in the bottom-right of the figure is not well populated.

\begin{figure}
    \centering
    \includegraphics[width=1\linewidth]{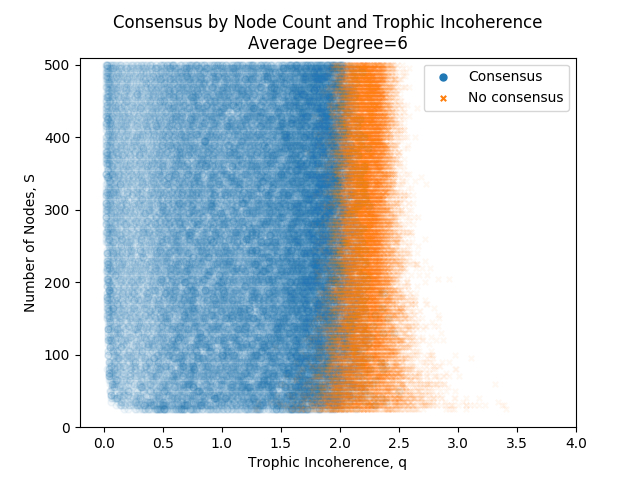}
    \caption{Each of 46,000 simulations is marked as a point, either reaching consensus or not. All networks have 5 basal nodes, an average out degree of 6, and varying number of nodes and trophic coherence.}
    \label{fig:consensus_regimes_nodes}
\end{figure}

An obvious question to ask is whether the size of a network has any effect on whether consensus is reached or not. Figure \ref{fig:consensus_regimes_nodes} shows simulations of majority rule dynamics on graphs with fixed average degree, $L/S=6$, fixed basal nodes, $B=5$, varying network size, $S$ and varying trophic incoherence parameter, $q$. With a fixed average degree $L/S=6$, the number of nodes doesn't seem to have a strong effect on whether consensus is reached or not, and the start of the area of no consensus is around $q=2$ for all network sizes, although it does appear to be slightly less at lower network sizes. 

\section*{Real Social Structures}

We mapped some real social structures onto graphs and measured their trophic coherence.

\subsection*{The Roman Army}

\begin{figure}
    \centering
    \includegraphics[width=1\linewidth]{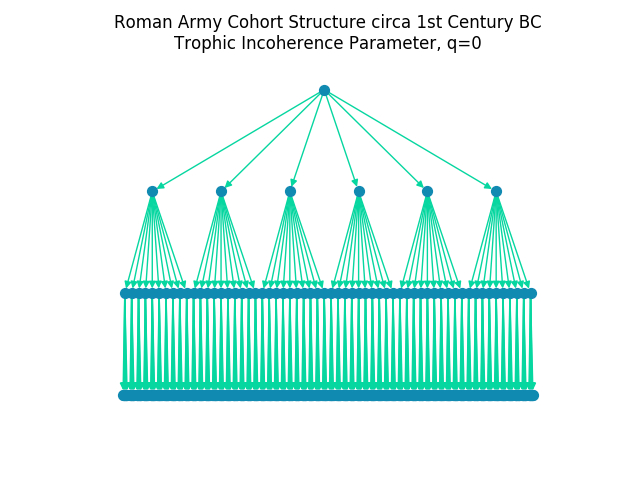}
    \caption{A cohort of the Roman Army from the 1st Century BC. Each approximately 480 man Cohort is made of 6 Centuria, each of which composed of 10 Conterbernia of 8 men \cite{Sage2013JanRomans}. The structure has perfect trophic coherence, with trophic incoherence parameter $q=0$.}
    \label{fig:romanArmy}
\end{figure}

An army is the prototypical hierarchical structure. Figure \ref{fig:romanArmy} shows the organisational structure of a Cohort in the Roman Army following the Marian reforms in the Late Roman Republic \cite{Sage2013JanRomans}. A legion was made up of 10 cohorts, a total of around 4500 men \cite{Sage2013JanRomans}. The Roman Army was incredibly effective, and employed a fighting style that required a high degree of coherent action between individuals and groups \cite{Sage2013JanRomans}. The highly coherent hierarchical structure with a large branching ratio was well suited to the demands and culture of the Roman Army, allowing orders from above to quickly reach all individuals.

\subsection*{The US Government}

\begin{figure}
    \centering
    \includegraphics[width=1\linewidth]{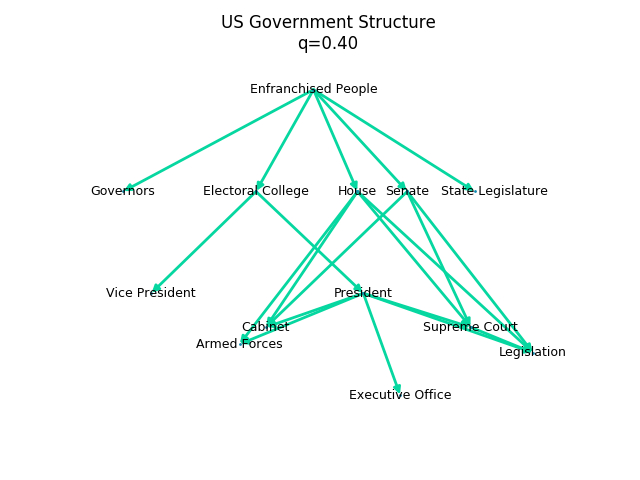}
    \caption{The general high-level structure of the US Government \cite{Beeman2010AugUSConan}. The Enfranchised People occupy the basal, leadership, node. The President's position does not stand out as particularly special in the network structure.}
    \label{fig:usGovt}
\end{figure}

The structure of the US Government is prescribed by the US Constitution, with an intention of creating a system of checks and balances that would prevent a demagogue from capturing too much power and becoming a tyrant \cite{Beeman2010AugUSConan}. Figure \ref{fig:usGovt} shows the high-level structure of the US Government, with a trophic incoherence parameter of $q=0.4$. This is interesting as the structure, as shown, falls into the regime of "tyranny", but with the Enfranchised People in a leadership position, and with the President not holding a particularly special position in the network. From this simple analysis, it would seem that the US Constitution successfully separates the powers in the US Government.  

The actual structure of the US Government is obviously much richer and more complicated. For example,  many of the nodes in this diagram could be expanded significantly. The Senate consists of 100 individual senators, there are 435 members of the House of Representatives \cite{Beeman2010AugUSConan}, and the Armed Forces and Judiciary have their own complex structures. And of course the Enfranchised People represents hundreds of millions of individuals. And there are many more complex and dynamic links of influence within the network. For example it would be reasonable to add an edge from Legislation to Enfranchised People, which would significantly change the trophic structure.

\subsection*{One 2 One Midwives}

\begin{figure}
    \centering
    \includegraphics[width=1\linewidth]{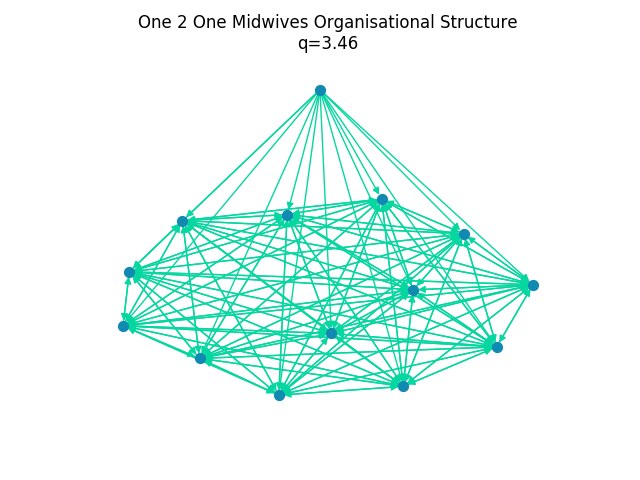}
    \caption{The structure of One 2 One Midwives, a Buurtzorg inspired healthcare organisation based in Liverpool, UK. The manager, top basal node, influences all midwives. All midwives influence each other.}
    \label{fig:midwives}
\end{figure}

The Buurtzorg healthcare model aims to improve patient care by embedding teams of around 12 healthcare workers within communities \cite{Buurtzorg}. The individual healthcare professionals are encouraged to help and support each other and are empowered to make autonomous decisions based on their professional training and patient knowledge \cite{Buurtzorg}. For a case study we look at One 2 One Midwives, a healthcare organisation inspired in part by the Buurtzorg healthcare model. One of the midwives was interviewed to discuss the organisational structure. Each of 13 midwives operate autonomously to care for a caseload of pregnant women, giving advice and support to each other. In addition a manager coordinates the midwives and can influence each of them. Figure \ref{fig:midwives} shows the organisational structure of One 2 One midwives, with trophic incoherence parameter, $q=3.46$. The real structure was reported by the interviewed midwife as more complex and dynamic, with different types of influence between manager to midwife and midwife to midwife. This kind of structure allows each midwife to operate autonomously, responding quickly and effectively to patient needs, while accessing support if needed. There is not a strong need for rapid consensus to influence from the manager (based on interview with H. Davey, August 6, 2019).

\section*{Discussion}

The trophic incoherence parameter, $q$, was shown to be useful in capturing the general topological coherence of directed graphs in a one-dimensional parameter. In the context of influence spreading in majority rule dynamics, three regimes of behaviour and topology were described:

\begin{enumerate}
  \item High trophic coherence, low q. Tree-like graphs with fast consensus. "Tyranny". 
  \item Medium trophic coherence, medium q. Some cycles and back edges with slow consensus. "Democracy". 
  \item Low trophic coherence, high q. Many cycles and back edges with no consensus. "Anarchy".
\end{enumerate}

For the low q regime, the dynamics of influence were successfully modelled as a series of Bernoulli trials with changing $p$, with a close fit between the Beta Negative Binomial distribution and the empirical distribution of time to consensus. 

In the medium q regime, the assumptions underpinning Bernoulli trials were stretched. The Beta Negative Binomial distribution was still a reasonable fit, but not as closely as in the low q regime. The presence of cycles and back edges was hypothesised as a reason for this poorer fit, and the slower consensus. 

In the high q regime, it was hypothesised that clique-like groups of nodes were blocking influence from spreading, with some nodes having more in-links from nodes of a higher trophic level than those of a lower trophic level. 

The location of the boundary between consensus and no consensus, in terms of the trophic incoherence parameter, $q$, was found to be largely independent of network size, given a fixed average degree.  While the number of edges was found to have an effect, with more connected graphs reaching the point of no consensus at lower values of the trophic incoherence parameter. 

Several real social structures were investigated with interesting insights, demonstrating that trophic coherence can be a useful tool in the description of real social systems. It is important to remember that there are myriad forms of social structures, each with their strengths and weaknesses. A very coherent structure that is good for the objectives of an army may be a poor choice for healthcare workers, and vice versa. Considering trophic coherence, it is striking that the current political status quo, in the West at least, is not an extreme but instead can be thought of as a balance between the efficiency of tyrannical hierarchy and the freedom of distributed anarchy.

The sociophysics model of majority rule dynamics is far too simple to accurately represent the actual behaviour of individuals within social systems, and we do not make that claim here. We have instead demonstrated one example of how the trophic incoherence parameter bounds regimes of vastly different network dynamics. We suggest that the consideration of the trophic coherence (or the related concept of non-normality) will provide significant insight in systems where the direction of interactions is important, as is the case in influence in social organisational systems. It is very early days in applying these concepts to social models, we encourage further work in this vein and our hope is that this will lead to useful tools in organisational and management science. 

The code developed during this study to calculate trophic levels and the trophic incoherence parameter has since being contributed to the open source python library networkx, and we direct any interested researchers to that resource.

\section*{Data Availability}
The datasets generated during and/or analysed during the current study are available from the corresponding author on reasonable request.

\section*{Additional information}
The authors declare no competing interests.

\bibliography{writeup}

\begin{thebibliography}{10}
\urlstyle{rm}
\expandafter\ifx\csname url\endcsname\relax
  \def\url#1{\texttt{#1}}\fi
\expandafter\ifx\csname urlprefix\endcsname\relax\def\urlprefix{URL }\fi
\expandafter\ifx\csname doiprefix\endcsname\relax\def\doiprefix{DOI: }\fi
\providecommand{\bibinfo}[2]{#2}
\providecommand{\eprint}[2][]{\url{#2}}

\bibitem{Sage2013JanRomans}
\bibinfo{author}{Sage, M.~M.}
\newblock \emph{\bibinfo{title}{{The Republican Roman Army: A Sourcebook
  (Routledge Sourcebooks for the Ancient World)}}}
  (\bibinfo{publisher}{limits}, \bibinfo{year}{2013}).

\bibitem{Laloux2014FebOrgs}
\bibinfo{author}{Laloux, F.}
\newblock \emph{\bibinfo{title}{{Reinventing Organizations: A Guide to Creating
  Organizations Inspired by the Next Stage in Human Consciousness}}}
  (\bibinfo{publisher}{Nelson Parker}, \bibinfo{year}{2014}).

\bibitem{Cohler1989Sep}
\bibinfo{author}{Cohler, A.~M.}
\newblock \emph{\bibinfo{title}{{Montesquieu: The Spirit of the Laws (Cambridge
  Texts in the History of Political Thought)}}} (\bibinfo{publisher}{Cambridge
  University Press}, \bibinfo{year}{1989}).

\bibitem{Beeman2010AugUSConan}
\bibinfo{author}{Beeman, R.}
\newblock \emph{\bibinfo{title}{{The Penguin Guide to the United States
  Constitution: A Fully Annotated Declaration of Independence, U.S.
  Constitution and Amendments, and Selections from the Federalist Papers}}}
  (\bibinfo{publisher}{Penguin Books}, \bibinfo{year}{2010}).

\bibitem{Franks2018MarAnarchy}
\bibinfo{author}{Franks, B.}, \bibinfo{author}{Jun, N.} \&
  \bibinfo{author}{Williams, L.}
\newblock \emph{\bibinfo{title}{{Anarchism: A Conceptual Approach (Routledge
  Studies in Radical History and Politics)}}} (\bibinfo{publisher}{Routledge},
  \bibinfo{year}{2018}).

\bibitem{Buurtzorg}
\bibinfo{title}{{The Buurtzorg Model - Buurtzorg International}}
  (\bibinfo{year}{2019}).
\newblock \bibinfo{note}{[Online; accessed 5. Sep. 2019]}.

\bibitem{Boccaletti2006Feb}
\bibinfo{author}{Boccaletti, S.}, \bibinfo{author}{Latora, V.},
  \bibinfo{author}{Moreno, Y.}, \bibinfo{author}{Chavez, M.} \&
  \bibinfo{author}{Hwang, D.-U.}
\newblock \bibinfo{journal}{\bibinfo{title}{{Complex networks: Structure and
  dynamics}}}.
\newblock {\emph{\JournalTitle{Phys. Rep.}}} \textbf{\bibinfo{volume}{424}},
  \bibinfo{pages}{175--308}, \doiprefix\url{10.1016/j.physrep.2005.10.009}
  (\bibinfo{year}{2006}).

\bibitem{Newman2003Mar}
\bibinfo{author}{Newman, M. E.~J.}
\newblock \bibinfo{journal}{\bibinfo{title}{{The Structure and Function of
  Complex Networks}}}.
\newblock {\emph{\JournalTitle{Comput. Phys. Commun.}}}
  \textbf{\bibinfo{volume}{147}}, \bibinfo{pages}{40--45},
  \doiprefix\url{10.1016/S0010-4655(02)00201-1} (\bibinfo{year}{2003}).

\bibitem{Albert2001Jun}
\bibinfo{author}{Albert, R.} \& \bibinfo{author}{Barabasi, A.-L.}
\newblock \bibinfo{journal}{\bibinfo{title}{{Statistical mechanics of complex
  networks}}}.
\newblock {\emph{\JournalTitle{arXiv}}}
  \doiprefix\url{10.1103/RevModPhys.74.47} (\bibinfo{year}{2001}).
\newblock \eprint{cond-mat/0106096}.

\bibitem{ParadoxMay1972Aug}
\bibinfo{author}{May, R.~M.}
\newblock \bibinfo{journal}{\bibinfo{title}{{Will a Large Complex System be
  Stable?}}}
\newblock {\emph{\JournalTitle{Nature}}} \textbf{\bibinfo{volume}{238}},
  \bibinfo{pages}{413--414}, \doiprefix\url{10.1038/238413a0}
  (\bibinfo{year}{1972}).

\bibitem{Johnson2014Dec}
\bibinfo{author}{Johnson, S.},
  \bibinfo{author}{Dom{\ifmmode\acute{\imath}\else\'{\i}\fi}nguez-Garc{\ifmmode\acute{\imath}\else\'{\i}\fi}a,
  V.}, \bibinfo{author}{Donetti, L.} \&
  \bibinfo{author}{Mu{\ifmmode\tilde{n}\else\~{n}\fi}oz, M.~A.}
\newblock \bibinfo{journal}{\bibinfo{title}{{Trophic coherence determines
  food-web stability}}}.
\newblock {\emph{\JournalTitle{Proc. Natl. Acad. Sci. U.S.A.}}}
  \textbf{\bibinfo{volume}{111}}, \bibinfo{pages}{17923--17928},
  \doiprefix\url{10.1073/pnas.1409077111} (\bibinfo{year}{2014}).

\bibitem{Olfati-Saber2004SepConsensusConnectivity}
\bibinfo{author}{Olfati-Saber, R.} \& \bibinfo{author}{Murray, R.~M.}
\newblock \bibinfo{journal}{\bibinfo{title}{{Consensus problems in networks of
  agents with switching topology and time-delays}}}.
\newblock {\emph{\JournalTitle{IEEE Trans. Autom. Control}}}
  \textbf{\bibinfo{volume}{49}} (\bibinfo{year}{2004}).

\bibitem{Shirley2005Sep}
\bibinfo{author}{Shirley, M. D.~F.} \& \bibinfo{author}{Rushton, S.~P.}
\newblock \bibinfo{journal}{\bibinfo{title}{{The impacts of network topology on
  disease spread}}}.
\newblock {\emph{\JournalTitle{Ecol. Complexity}}}
  \textbf{\bibinfo{volume}{2}}, \bibinfo{pages}{287--299},
  \doiprefix\url{10.1016/j.ecocom.2005.04.005} (\bibinfo{year}{2005}).

\bibitem{Cohen2009}
\bibinfo{author}{Cohen, R.} \& \bibinfo{author}{Havlin, S.}
\newblock \emph{\bibinfo{title}{{Percolation in Complex Networks}}}
  (\bibinfo{publisher}{Springer, New York, NY}, \bibinfo{year}{2009}).

\bibitem{Guo16}
\bibinfo{author}{Guo, W.} \& \bibinfo{author}{Lu, X.}
\newblock \bibinfo{journal}{\bibinfo{title}{{Core identification and attack
  strategies against regenerative complex networks}}}.
\newblock {\emph{\JournalTitle{IET Electronics Letters}}}
  \textbf{\bibinfo{volume}{52}} (\bibinfo{year}{2016}).

\bibitem{Pagani19RSOS}
\bibinfo{author}{Pagani, A.} \emph{et~al.}
\newblock \bibinfo{journal}{\bibinfo{title}{Resilience or robustness:
  identifying topological vulnerabilities in rail networks}}.
\newblock {\emph{\JournalTitle{Royal Society Open Science}}}
  \textbf{\bibinfo{volume}{6}}, \bibinfo{pages}{181301},
  \doiprefix\url{10.1098/rsos.181301} (\bibinfo{year}{2019}).

\bibitem{Barzel13}
\bibinfo{author}{Barzel, B.} \& \bibinfo{author}{Barabasi, A.}
\newblock \bibinfo{journal}{\bibinfo{title}{{Universality of Network
  Dynamics}}}.
\newblock {\emph{\JournalTitle{Nature Physics}}} \textbf{\bibinfo{volume}{9}}
  (\bibinfo{year}{2013}).

\bibitem{Gao16}
\bibinfo{author}{Gao, J.}, \bibinfo{author}{Barzel, B.} \&
  \bibinfo{author}{Barabasi, A.}
\newblock \bibinfo{journal}{\bibinfo{title}{{Universal resilience patterns in
  complex networks}}}.
\newblock {\emph{\JournalTitle{Nature}}} \textbf{\bibinfo{volume}{530}}
  (\bibinfo{year}{2016}).

\bibitem{Asllani2017Jun}
\bibinfo{author}{Asllani, M.} \& \bibinfo{author}{Carletti, T.}
\newblock \bibinfo{journal}{\bibinfo{title}{{Topological resilience in
  non-normal networked systems}}}.
\newblock {\emph{\JournalTitle{arXiv}}}
  \doiprefix\url{10.1103/PhysRevE.97.042302} (\bibinfo{year}{2017}).
\newblock \eprint{1706.02703}.

\bibitem{Johnson2019Aug}
\bibinfo{author}{Johnson, S.}
\newblock \bibinfo{journal}{\bibinfo{title}{{Digraphs are different: Why
  directionality matters in complex systems}}}.
\newblock {\emph{\JournalTitle{arXiv}}}  (\bibinfo{year}{2019}).
\newblock \eprint{1908.07025}.

\bibitem{Klaise2016JunGPPM}
\bibinfo{author}{Klaise, J.} \& \bibinfo{author}{Johnson, S.}
\newblock \bibinfo{journal}{\bibinfo{title}{{From neurons to epidemics: How
  trophic coherence affects spreading processes}}}.
\newblock {\emph{\JournalTitle{Chaos: An Interdisciplinary Journal of Nonlinear
  Science}}} \textbf{\bibinfo{volume}{26}}, \bibinfo{pages}{065310},
  \doiprefix\url{10.1063/1.4953160} (\bibinfo{year}{2016}).

\bibitem{Asllani2018Dec}
\bibinfo{author}{Asllani, M.}, \bibinfo{author}{Lambiotte, R.} \&
  \bibinfo{author}{Carletti, T.}
\newblock \bibinfo{journal}{\bibinfo{title}{{Structure and dynamical behavior
  of non-normal networks}}}.
\newblock {\emph{\JournalTitle{Sci. Adv.}}} \textbf{\bibinfo{volume}{4}},
  \bibinfo{pages}{eaau9403}, \doiprefix\url{10.1126/sciadv.aau9403}
  (\bibinfo{year}{2018}).

\bibitem{sociophysicsCastellano2007Oct}
\bibinfo{author}{Castellano, C.}, \bibinfo{author}{Fortunato, S.} \&
  \bibinfo{author}{Loreto, V.}
\newblock \bibinfo{journal}{\bibinfo{title}{{Statistical physics of social
  dynamics}}}.
\newblock {\emph{\JournalTitle{arXiv}}}
  \doiprefix\url{10.1103/RevModPhys.81.591} (\bibinfo{year}{2007}).
\newblock \eprint{0710.3256}.

\bibitem{Brauer2008}
\bibinfo{author}{Brauer, F.}
\newblock \bibinfo{journal}{\bibinfo{title}{{Compartmental Models in
  Epidemiology}}}.
\newblock {\emph{\JournalTitle{SpringerLink}}} \bibinfo{pages}{19--79},
  \doiprefix\url{10.1007/978-3-540-78911-6_2} (\bibinfo{year}{2008}).

\bibitem{Motter2002Dec}
\bibinfo{author}{Motter, A.~E.} \& \bibinfo{author}{Lai, Y.-C.}
\newblock \bibinfo{journal}{\bibinfo{title}{{Cascade-based attacks on complex
  networks}}}.
\newblock {\emph{\JournalTitle{Phys. Rev. E}}} \textbf{\bibinfo{volume}{66}},
  \bibinfo{pages}{065102(R)}, \doiprefix\url{10.1103/PhysRevE.66.065102}
  (\bibinfo{year}{2002}).

\bibitem{LooplessnessJohnson2017May}
\bibinfo{author}{Johnson, S.} \& \bibinfo{author}{Jones, N.~S.}
\newblock \bibinfo{journal}{\bibinfo{title}{{Looplessness in networks is linked
  to trophic coherence}}}.
\newblock {\emph{\JournalTitle{Proc. Natl. Acad. Sci. U.S.A.}}}
  \bibinfo{pages}{201613786}, \doiprefix\url{10.1073/pnas.1613786114}
  (\bibinfo{year}{2017}).

\bibitem{Levine1980MarTrophicLevels}
\bibinfo{author}{Levine, S.}
\newblock \bibinfo{journal}{\bibinfo{title}{{Several measures of trophic
  structure applicable to complex food webs}}}.
\newblock {\emph{\JournalTitle{J. Theor. Biol.}}}
  \textbf{\bibinfo{volume}{83}}, \bibinfo{pages}{195--207},
  \doiprefix\url{10.1016/0022-5193(80)90288-X} (\bibinfo{year}{1980}).

\bibitem{Pagani19ASCE}
\bibinfo{author}{Pagani, A.}, \bibinfo{author}{Meng, F.}, \bibinfo{author}{Fu,
  G.}, \bibinfo{author}{Musolesi, M.} \& \bibinfo{author}{Guo, W.}
\newblock \bibinfo{journal}{\bibinfo{title}{Quantifying networked resilience
  via multi-scale feedback loops in water distribution networks}}.
\newblock {\emph{\JournalTitle{ASCE Water Resources Planning Management (to
  appear)}}}  (\bibinfo{year}{2019}).

\bibitem{Moutsinas2019Aug}
\bibinfo{author}{Moutsinas, G.}, \bibinfo{author}{Shuaib, C.},
  \bibinfo{author}{Guo, W.} \& \bibinfo{author}{Jarvis, S.}
\newblock \bibinfo{journal}{\bibinfo{title}{{Generalised trophic levels and
  graph hierarchy}}}.
\newblock {\emph{\JournalTitle{ResearchGate}}}  (\bibinfo{year}{2019}).

\bibitem{Jacquet2016Aug}
\bibinfo{author}{Jacquet, C.} \emph{et~al.}
\newblock \bibinfo{journal}{\bibinfo{title}{{No
  complexity{\textendash}stability relationship in empirical ecosystems}}}.
\newblock {\emph{\JournalTitle{Nat. Commun.}}} \textbf{\bibinfo{volume}{7}},
  \bibinfo{pages}{12573}, \doiprefix\url{10.1038/ncomms12573}
  (\bibinfo{year}{2016}).

\bibitem{MotifsKlaise2017Nov}
\bibinfo{author}{Klaise, J.} \& \bibinfo{author}{Johnson, S.}
\newblock \bibinfo{journal}{\bibinfo{title}{{The origin of motif families in
  food webs}}}.
\newblock {\emph{\JournalTitle{Sci. Rep.}}} \textbf{\bibinfo{volume}{7}},
  \bibinfo{pages}{16197--11}, \doiprefix\url{10.1038/s41598-017-15496-1}
  (\bibinfo{year}{2017}).

\bibitem{Bollobas1998}
\bibinfo{author}{Bollob{\ifmmode\acute{a}\else\'{a}\fi}s, B.}
\newblock \bibinfo{journal}{\bibinfo{title}{{Random Graphs}}}.
\newblock {\emph{\JournalTitle{SpringerLink}}} \bibinfo{pages}{215--252},
  \doiprefix\url{10.1007/978-1-4612-0619-4_7} (\bibinfo{year}{1998}).

\bibitem{erdos59a}
\bibinfo{author}{Erd\"{o}s, P.} \& \bibinfo{author}{R\'{e}nyi, A.}
\newblock \bibinfo{journal}{\bibinfo{title}{On random graphs i}}.
\newblock {\emph{\JournalTitle{Publicationes Mathematicae Debrecen}}}
  \textbf{\bibinfo{volume}{6}}, \bibinfo{pages}{290} (\bibinfo{year}{1959}).

\bibitem{Grinstead2012Oct}
\bibinfo{author}{Grinstead, C.~M.} \& \bibinfo{author}{Snell, J.~L.}
\newblock \emph{\bibinfo{title}{{Introduction to Probability: Second Revised
  Edition}}} (\bibinfo{publisher}{American Mathematical Society},
  \bibinfo{year}{2012}).

\bibitem{Johnson2005SepBetaNegBinomial}
\bibinfo{author}{Johnson, N.~L.}, \bibinfo{author}{Kotz, S.} \&
  \bibinfo{author}{Kemp, A.~W.}
\newblock \emph{\bibinfo{title}{{Univariate Discrete Distributions, 3rd Edition
  (Wiley Series in Probability and Statistics)}}} (\bibinfo{publisher}{John
  Wiley {\&} Sons Inc}, \bibinfo{year}{2005}).

\end{thebibliography}

\section*{Author Contributions}
W.G. and S.J. conceived of the presented idea. C.P. developed the theory and performed the simulations.  W.G. and S.J. supervised throughout. C.P. took the lead in writing the manuscript with input and guidance from W.G..

\section*{Supplementary Information}

\section{Trophic Levels}

Considering the food web network adjacency matrix, $A$, such that $a_{ij}$ is the amount of biomass that species $j$ predates from species $i$. The trophic level of a species, $s_j$, is defined as the weighted mean of the trophic levels of the species that $j$ preys upon, plus one \cite{Johnson2014Dec}.

\begin{equation}
s_j = \frac{\sum_{i} a_{ij} s_i}{\sum_{i} a_{ij}} + 1 \label{eqn:trophic_level_definition2}
\end{equation}

This can be simplified by considering the weighted biomass transfer along edges, $p_{ij}$, found by normalising the biomass transfer along edges to each predator species,

\begin{equation}
p_{ij} = \frac{a_{ij}}{\sum_j a_{ij}} \label{eqn:weighted_biomass}
\end{equation}

So that equation \ref{eqn:trophic_level_definition2} becomes

\begin{equation}
s_j - \sum_{i} p_{ij} s_i = 1 \label{eqn:trophic_level_definition_weighted}
\end{equation}

Equation \ref{eqn:trophic_level_definition_weighted} describes a linear system.

\begin{equation}
\begin{bmatrix}
1 & - p_{21} & -p_{31} & \dots & -p_{n1} \\
-p_{12} & 1 & -p_{32} & \dots & -p_{n2} \\
\dots  & \dots  & \dots  & \dots & \dots  \\
-p_{1n} & -p_{2n} & -p_{3n} & \dots & 1 
\end{bmatrix}
\begin{bmatrix}
s_1 \\ s_2 \\ \dots \\ s_n 
\end{bmatrix}
=
\begin{bmatrix}
1 \\ 1 \\ \dots \\ 1
\end{bmatrix}
\end{equation}

This can be written in matrix notation, where P is the matrix with entries $p_{ij}$ and $\mathbf{1}$ is a vector of 1s. 

\begin{equation}
 \mathbf{s} = ((I-P)^{-1})^{T}  \mathbf{1}  \label{eqn:trophiclevels} 
\end{equation}

This linear system can be solved with a unique solution with the sufficient condition that there is at least one basal species that does not prey on any other species. 

Equation \ref{eqn:trophiclevels} can be arrived at through consideration of the linear system. Alternatively, the trophic level can be defined in terms of how many levels biomass travels through to get to each species \cite{Levine1980MarTrophicLevels}, by considering the biomass weighted adjacency matrix, $P$. 

\begin{equation}
s_j = 1 + \sum_{i} p_{i,j} + \sum_{i} p_{i,j}^{(2)} + .. + \sum_{i} p_{i,j}^{(N)}
\end{equation}

Where N is the maximum number of steps in a food web chain that we wish to consider. This can be written as

\begin{equation}
 \mathbf{s} = (I + P + P^2 + .. P^N)^T \mathbf{1} \label{eqn:trophiclevels2} 
\end{equation}

Comparing this to equation \ref{eqn:trophiclevels}, it is clear that equation \ref{eqn:trophiclevels2} is equivalent if 

\begin{equation}
(I + P + P^2 + .. P^N) = (I-P)^{-1}
\end{equation}

Multiplying both sides by $(I-P)$, we see that this is equivalent if $P^{N} \to 0$ as $N \to \infty$. Each basal species, by definition, has in-degree zero and so that the associated column in $P$ must be zeroes. The matrix elements, $p_{ij}$, were defined as being normalised along $j$ so that the non-zero columns of $P$ sum to 1. Give the condition that each node has at least one path form a basal node, we can see that this is sufficient for $P^{N} \to 0$ as $N \to \infty$, and so equations \ref{eqn:trophiclevels} and \ref{eqn:trophiclevels2} are equivalent.

\section{Calculation of Cycles in a Graph}

If we consider the squared adjacency matrix of a directed graph, $A^2$. The elements of this adjacency matrix can be written as:

\begin{equation}
a^{(2)}_{ij} = \sum_{k=1}^S a_{ik}a_{kj}
\end{equation}

That is, $a^{(2)}_{ij}$ gives the number of paths of length 2 in the graph from node $i$ to $j$. Through induction, it can be seen that $a^{(n)}_{ij}$ gives the number of paths of length n in the graph from node $i$ to $j$. The diagonal entries of $A^n$, $a_{ii}^{n}$, will therefore give the number of paths of length $n$ that begin and end at node $i$, which is the definition of a cycle of length $n$. In order to find the total number of cycles of length $n_{cycle}$ in a given graph we can sum the diagonals of the graph's adjacency matrix raised to that power, given by the Trace.

\begin{equation}
\text{cycle count} = Tr(A^{n_{cycle}}) \label{eqn_cycle_counts}
\end{equation}

Equation \ref{eqn_cycle_counts} was used to calculate the cycle counts in Figure \ref{fig:trophic_incoherence}(a).

\section{Derivation for Cycles}

\subsection*{Pure cycle}

We will begin by considering influence dynamics in a pure cycle, with no basal node. In a pure cycle, all nodes have a single parent node, and so when the majority rule algorithm randomly selects a node it takes on the state of its sole parent. A node will switch state if and only if it is in the opposite state of its parent node. Through symmetry considerations, there must be the same number of nodes ready to switch from state 1 to -1 if selected as vice versa. (The number of "fringes" between blocks of 1 and -1 nodes must be the same on a circle). At each timestep, the  transition probabilities of $v(t)$ are therefore:

\begin{equation}
P_{pure}(v \to v + 1) = \frac{p_f}{2}\label{eqn_pure_plus}
\end{equation}
\begin{equation}
P_{pure}(v \to v - 1) = \frac{p_f}{2}\label{eqn_pure_minus}
\end{equation}
\begin{equation}
P_{pure}(v \to v) = 1-p_f \label{eqn_pure_equal}
\end{equation}

Where $p_f$ is the probability of selecting a child node at a "fringe" where the parent has a different state to the child. There are absorbing states at $v(t)=0$ and $v(t)=n$, where all nodes are in the same state and there are no fringes. There is some path dependency involved in $p_f(v)$, as certain configurations can only be reached by other configurations. The form of $p_f(v)$ is not necessary for our analysis. From the symmetry between equations \ref{eqn_pure_plus} and \ref{eqn_pure_minus} we can see that

\begin{equation}
<v(t)> = v(0)
\end{equation}

\subsection*{Cycle with a basal node}
\begin{figure}
    \centering
    \includegraphics[width=1\linewidth]{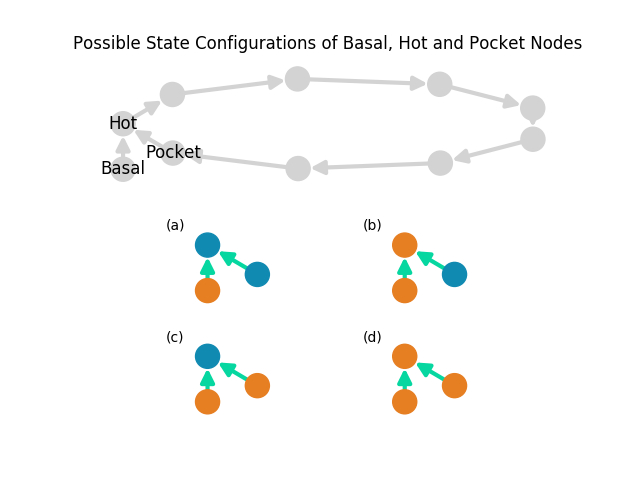}
    \caption{The basal node is always in state -1. The hot and pocket nodes can be in one of 4 configurations of states. Orange nodes here are in state -1, the same as the basal node. Blue nodes are in state 1, the initial state of nodes in the cycle.}
    \label{fig:cycle_configurations}
\end{figure}

Now, we add one basal node that influences one of the nodes within the cycle, which we will call the "hot" node. The other important node to consider is the parent of the hot node within the cycle, which we will call the "pocket" node (See Figure \ref{fig:cycle_configurations}). We will consider how adding the basal node changes the dynamics from the pure cycle. At each timestep, the only node effected by the basal node is the hot node, and so we can consider the dynamics in the rest of the cycle as proceeding as before.

The basal node is always in state $x_{basal}= -1$. This gives 4 possible configurations for states of the hot and pocket nodes, as shown in Figure \ref{fig:cycle_configurations}. For each of these configurations we will write down how the transition probabilities of $v(t)$ differ compared to the pure cycle. 

\paragraph{Configuration a. $x_{hot}=1, x_{pocket}=1$}. 
Given that the hot node is selected by the algorithm, and that the system begins in configuration a, there is a 50$\%$ chance that the basal node will influence the hot node.

\begin{equation}
    P_{basal}(v \to v|hot,a)=\frac{1}{2} \label{eqn_conf_start}    
\end{equation}
\begin{equation}
  P_{basal}(v \to v-1|hot,a)=\frac{1}{2}  
\end{equation}

Comparing this to the pure cycle in configuration a (without a basal node), given that the hot node is selected by the algorithm, there is no chance of any change in $v(t)$

\begin{equation}
P_{pure}(v \to v|hot,a)=1
\end{equation}

\paragraph{Configuration b. $x_{hot}=-1, x_{pocket}=1$}. 
Following similar considerations as in Configuration a, we can see that

\begin{equation}
  P_{basal}(v \to v|hot,b)=\frac{1}{2}  
\end{equation}
\begin{equation}
  P_{basal}(v \to v+1|hot,b)=\frac{1}{2}  
\end{equation}
\begin{equation}
  P_{pure}(v \to v+1|hot,b)=1  
\end{equation}

\paragraph{Configuration c. $x_{hot}=1, x_{pocket}=-1$}.
The pocket and basal node agree, so there is no change from the dynamics of the pure cycle.

\begin{equation}
  P_{pure}(v \to v+1|hot,c) = P_{basal}(v \to v+1|hot,c)=1  
\end{equation}

\paragraph{Configuration d. $x_{hot}=-1, x_{pocket}=-1$}.
Again, there is no change from the dynamics of the pure cycle.

\begin{equation}
  P_{pure}(v \to v|hot,d) = P_{basal}(v \to v|hot,d)=1  \label{eqn_conf_end}
\end{equation}

Considering equations \ref{eqn_conf_start}-\ref{eqn_conf_end}, we can adapt the pure cycle transition probability equations \ref{eqn_pure_plus}, \ref{eqn_pure_minus} and \ref{eqn_pure_equal} to the basal cycle, in terms of the probability of selecting the "hot" node, $P(hot)$ and the probability of being in configuration $m$, $P(m)$:

\begin{equation}
    P_{basal}(v \to v + 1) = \frac{p_f}{2} -\frac{1}{2}P(hot)P(b) \label{eqn_basal_plus}
\end{equation}
\begin{equation}
  P_{basal}(v \to v - 1) = \frac{p_f}{2} + \frac{1}{2}P(hot)P(a)  \label{eqn_basal_minus}
\end{equation}
\begin{equation}
  P_{basal}(v \to v) = 1-p_f + \frac{1}{2}P(hot)P(b) - \frac{1}{2}P(hot)P(a)
\end{equation}

Assuming large n, we can write the change in the expected value of $v$ as: 

\begin{equation}
  \frac{d<v>}{dt} \propto P(v \to v+1) - P(v \to v-1)  
\end{equation}

Substituting in equations \ref{eqn_basal_plus} and \ref{eqn_basal_minus}, and adding a constant of proportionality, $k$,

\begin{equation}
  \frac{d<v>}{dt} = - k \left(\frac{1}{2}P(hot)P(a) + \frac{1}{2}P(hot)P(b)\right)
\end{equation}

\begin{equation}
  \frac{d<v>}{dt} = - k \frac{1}{2}P(hot)P(a \cup b)
\end{equation}

The probability of the algorithm selecting the hot node, $P(hot)=\frac{1}{n}$. Configurations $a$ and $b$ are where the pocket node disagrees with the basal node (Figure \ref{fig:cycle_configurations}). The probability that the basal and pocket nodes disagree is equal to $P(a \cup b)=\frac{v}{n}$. Substituting these probabilities in,

\begin{equation}
  \frac{d<v>}{dt} = - k \frac{<v>}{2n^2}
\end{equation}

This can be solved to find that

\begin{equation}
  <v(t)> = A e^{-\frac{kt}{2n^2}} \label{eqn_cycle_exp}  
\end{equation}

Considering our initial conditions of the basal node being in state -1 while all nodes in the cycle are in state 1:

\begin{equation}
    A = <v(0)> = n-1
\end{equation}

Equation \ref{eqn_cycle_exp} describes exponential decay. We can replace the constant $k$ by considering the half-life of the decay, in this case how long we expect it to take for half of the nodes in the cycle to be influenced by the basal node through the mechanisms described in equations \ref{eqn_basal_plus} and \ref{eqn_basal_minus}. The hot node will be selected every $n$ timesteps. In $n-1$ selections of the hot node, we expect, on average, $v$ opportunities for the basal node to change the mechanics from the pure cycle mechanism, of which it will succeed half of the time. The expected half life is therefore $n(n-1)$ timesteps.

\begin{equation}
    <v(t)> = (n-1)e^{-\frac{ln2}{n(n-1)}t}
\end{equation}

\end{document}